\documentclass[a4paper,10pt,twoside]{article}

\usepackage{definitions/CMST}

\newcommand\addTitle{Stability Analysis of Three Coupled Kerr Oscillators: Implications for Quantum Computing}
\newcommand\addShortTitle{Stability Analysis of Three Coupled Kerr Oscillators}

\newcommand\addAuthors{K. Chmielewski$^1$, K. Grygiel$^{2*}$, K. Bartkiewicz$^{2}$}
\newcommand\addAuthorsNoSuperscript{K. Chmielewski, K. Grygiel, K. Bartkiewicz}
\newcommand\addAffiliations{

$^1$ Faculty of Physics and Astronomy, Adam Mickiewicz University,
61-614 Pozna\'n, Poland\\

$^2$ Department of Quantum Information, Faculty of Physics and
Astronomy, Adam Mickiewicz University,
61-614 Pozna\'n, Poland\\
$^*$E-mail: grygielk@amu.edu.pl (corresp.)}

\newcommand\addAbstract{
We investigate the classical dynamics of optical nonlinear Kerr
couplers, focusing on their potential relevance to quantum
computing applications. The system consists of three Kerr-type
nonlinear oscillators arranged in two configurations: a triangular
arrangement, where each oscillator is coupled to the others, and a
sandwich arrangement, where only the middle oscillator interacts
with the two outer ones. The system is driven by an external
periodic field and includes dissipative processes. Its evolution
is governed by six non-autonomous differential equations derived
from a Kerr Hamiltonian with nonlinear coupling terms. We show
that even for identical Kerr media, the interplay between
nonlinear couplings and mismatched fundamental and pump
frequencies leads to rich and complex dynamics, including the
emergence of multiple stable attractors. These attractors are
highly sensitive to both the coupling configuration and initial
conditions. A key contribution of this work is a detailed
stability analysis based on the numerical calculation of Lyapunov
exponents, which reveals transitions from regular to chaotic
dynamics as damping is reduced. We identify critical damping
thresholds for the onset of chaos and characterize phenomena such
as chaotic beats. These results offer insights for potential
experimental realizations and are directly relevant to emerging
quantum technologies, where Kerr parametric oscillators play a
central role in quantum gates, error correction protocols, and
quantum neural network architectures. }

\newcommand\addKeywords{
Kerr oscillators, Lyapunov exponents, quantum technology, chaotic
beats}

\begin{document}
\maketitle
\begin{multicols}{2}
\section{Introduction}
Nonlinear optical systems based on the Kerr effect have emerged as
a key area of overlap between classical nonlinear dynamics and
quantum information science. Extensive research on such systems
has explored both classical \cite{BoydBook, KivsharBook} and
quantum \cite{ScullyBook, GerryBook, WallsBook, Gu2017} properties
of the generated optical fields, offering valuable insights into
fundamental physical processes and enabling a range of practical
applications (as briefly reviewed in Sec. II). In particular,
systems of coupled oscillators with Kerr-type nonlinearity have
garnered significant interest due to their rich dynamical behavior
and potential uses in optical signal processing, all-optical
switching, and, more recently, quantum information processing.

The nonlinear Kerr effect, defined by an intensity-dependent
refractive index, facilitates self-phase modulation and enables
complex interactions when multiple optical fields are coupled. In
systems of coupled Kerr oscillators, this nonlinearity gives rise
to a wide spectrum of dynamical behaviors, ranging from regular
periodic motion to intricate chaotic dynamics. Such richness makes
these systems valuable both for advancing theoretical
understanding and for enabling diverse technological applications.
Recent progress in nanophotonics and integrated optics has
significantly increased the feasibility of experimentally
realizing these systems (see \cite{Cui2022} and references
therein). Notably, silicon nitride microresonators have
demonstrated high-efficiency optical parametric oscillation,
achieving conversion efficiencies as high as 29\%
\cite{Perez2022}.

A particularly significant development is the recent emergence of
Kerr parametric oscillators (KPOs) as promising building blocks
for quantum processors, especially in superconducting circuit
platforms \cite{Gu2017, Yin2021,IBM2023}. Over the past few years
(2022--2025), these systems have demonstrated key advantages,
including high gate fidelities, enhanced error resilience, and
increased computational capabilities \cite{Kanao2022}. The
extension from two to three coupled oscillators---the central
focus of this work---marks a critical threshold, enabling quantum
functionalities that are unattainable in simpler configurations.
Three-oscillator systems exhibit significantly richer phase-space
structures, supporting up to eight stable fixed points under
specific parameter regimes \cite{Margiani2025}, thereby offering
an expanded state space for quantum information encoding and
manipulation.

The mathematical description of three coupled KPOs in the quantum
regime involves a Hamiltonian comprising multiple nonlinear
interaction terms:
\begin{eqnarray}
H &=& \sum_{i=1}^{3} \left[ \Delta_i a_i^{\dagger}a_i -
\frac{K_i}{2}(a_i^{\dagger}a_i)^2 + \frac{p_i}{2}(a_i^2 +
{a_i^{\dagger}}^2) \right]
\nonumber \\
&&+ \sum_{i,j>i}^{3} J_{ij}(a_i^{\dagger}a_j + a_i a_j^{\dagger}),
\end{eqnarray}
where $\Delta_i$ denotes the detuning frequency, $K_i$ the Kerr
nonlinearity strength, $p_i$ the pumping amplitude, and $J_{ij}$
the coupling strength between oscillators. While this Hamiltonian
formulation applies to the few-photon quantum regime, our
classical analysis offers complementary insights into the system's
behavior in higher-excitation regimes, especially concerning
stability, multistability, and the onset of chaos.

In the commercial sector, quantum computing platforms based on
coupled nonlinear oscillators are beginning to take shape. A
notable example is IBM's Quantum System Two, launched in 2023,
which introduced the first modular quantum computer featuring
three coupled Heron processors. This architecture supports the
execution of up to 1800 quantum gates within coherence
times---nearly quadrupling the capacity of earlier systems
\cite{IBM2023}. Such advances emphasize the growing importance of
a detailed understanding of multi-oscillator Kerr systems, both
from theoretical and practical standpoints.

Earlier work by \'Sliwa and Grygiel \cite{Sliwa2012} explored the
dynamics of two coupled Kerr oscillators, uncovering rich
phase-space structures characterized by multiple coexisting
attractors and transitions between regular and chaotic behavior.
Building upon their findings, the present study extends the
framework to three coupled oscillators, introducing additional
degrees of freedom and novel coupling topologies. These extensions
give rise to significantly more intricate dynamical behavior,
enabling the exploration of new stability regimes and potential
applications in quantum information processing.

It is also worth highlighting previous fully quantum-mechanical
treatments of triple Kerr oscillator couplers. Notably, Kalaga et
al. \cite{Kalaga2016} demonstrated that such a system can function
as a nonlinear quantum scissors device and effectively operate as
a three-qubit model. More recently, Hanapi et al.
\cite{Hanapi2024} investigated an optical coupler composed of
three second-harmonic generation systems, focusing on the
generation of nonclassical optical fields. However, these studies
primarily concentrated on quantum aspects. In contrast, our
present work emphasizes the classical regime, particularly through
a detailed stability analysis using Lyapunov exponents---an
approach that, to our knowledge, has not yet been applied to
systems of three coupled Kerr oscillators.

The study of chaotic dynamics in nonlinear optical systems holds
both fundamental and practical importance. On a fundamental level,
it sheds light on the mechanisms governing the transition from
regular to chaotic behavior in complex nonlinear systems. From a
practical perspective, controlled chaos has been proposed for
diverse applications, including secure optical communication
\cite{VanWiggeren1998}, high-speed random number generation
\cite{Uchida2008}, and photonic reservoir computing
\cite{Brunner2013}.

This paper builds upon previous research by systematically
analyzing the dynamics and stability of three coupled Kerr
oscillators under different coupling configurations. Our primary
focus is on the influence of nonlinear couplings on system
stability and phase-space evolution, with particular attention to
the onset of chaos as key parameters are varied.

\begin{figure}[H]
\begin{center}
\includegraphics[width=5cm]{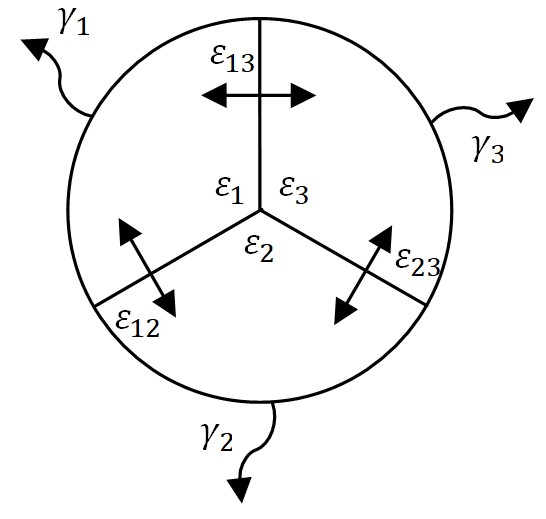}

\vspace{5mm}

\includegraphics[width=7cm]{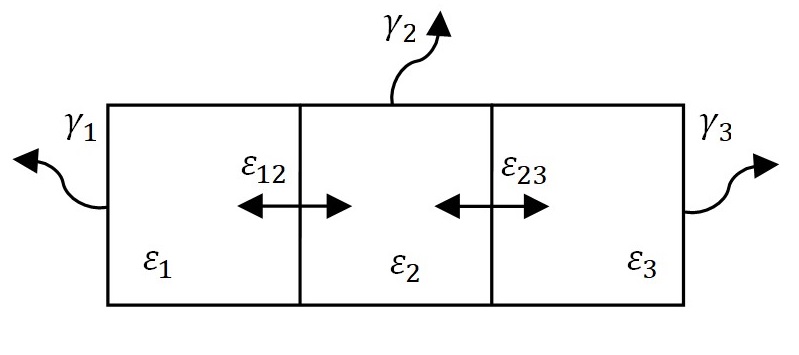}
\end{center}
\caption{Cross-sections of two coupling configurations: the
triangular arrangement (top), in which all three oscillators are
mutually coupled, and the sandwich arrangement (bottom), where the
middle oscillator is coupled to both others, but the first and
last oscillators have no direct coupling.} \label{fig:fig1}
\end{figure}

The novel contributions of our study include: (1) a comparative
analysis of both triangular and sandwich coupling topologies (Fig. \ref{fig:fig1}); (2)
a comprehensive Lyapunov exponent analysis identifying transitions
to chaotic regimes; (3) the discovery of multiple coexisting stable attractors with distinct phase-space structures dependent on coupling parameters (as well as mapping their basins of attraction); and (4) the
demonstration of signal patterns known as chaotic beats, generated
by the system under specific conditions.

Before introducing our model and showing its numerical solutions,
we first highlight the significance of the optical Kerr effect in
contemporary quantum nonlinear physics.

\section{On the Fundamental and Practical Role of the Optical Kerr Effect}

Although this paper focuses on the classical Kerr model, the
underlying phenomena are highly relevant for quantum technologies
due to their fundamental importance and wide range of
applications.

The optical Kerr effect plays a central role in quantum state
engineering and quantum information processing (QIP), and has
attracted sustained interest due to its rich nonlinear dynamics
and broad applicability across various platforms in quantum optics
and related fields. These include cavity quantum electrodynamics
(QED)~\cite{GerryBook}, circuit QED (based on superconducting
quantum circuits coupled to microwave resonators)~\cite{Gu2017,
Yin2021}, atom optics (using Rydberg atoms, cold atomic gases, and
Bose-Einstein condensates)~\cite{HarocheBook}, as well as cavity
optomechanical systems (see, e.g.,~\cite{Aldana2013, Wang2019}).
Moreover, Kerr-type systems provide prototypical models for
exploring chaotic dynamics and nonlinear quantum control, which
are the central focus of this paper.

A qubit---whether a natural or artificial atom such as a
superconducting circuit---dispersively coupled to a resonator
(like transmission line resonator) provides a versatile platform
for exploring Kerr-type light-matter interactions. In this
dispersive limit, the qubit induces measurable frequency and phase
shifts in the resonator's spectrum, enabling effective Kerr
nonlinearities. This approach is widely employed to realize
Kerr-type interactions, particularly when the qubit-resonator
(i.e., light-matter) coupling reaches the strong, ultrastrong, or
even deep-strong regimes~\cite{Kockum2019, Qin2024}.

The Kerr effect, which induces an intensity-dependent refractive
index, gives rise to a variety of nonlinear optical phenomena such
as dispersive optical bistability \cite{WallsBook}, self-focusing
and self-phase modulation \cite{Tanas1996}. It also plays a
central role in quantum light control by enabling photon
blockade---an effect that suppresses the absorption of multiple
photons, allowing for the generation of single
photons~\cite{Leonski1994, Imamoglu1997}. This effect has been
demonstrated in numerous experiments~(see~\cite{Gu2017,
Leonski2011} for references). Beyond single-photon blockade, the
Kerr effect enables a wide range of advanced phenomena, including
multi-photon blockade~\cite{Miranowicz2013, Hamsen2017,
Kowalewska2019}, nonreciprocal~\cite{Huang2018} and
chiral~\cite{Zuo2024} photon blockade effects, as well as phonon
blockade, in which mechanical excitations (phonons) are
suppressed~\cite{Liu2010}, and hybrid photon-phonon
blockade~\cite{Abo2022}. Photon~\cite{Leonski2004, Miranowicz2006}
and phonon~\cite{Miranowicz2016} blockade effects in coupled Kerr
oscillator systems have also been studied as mechanisms for
generating maximally entangled states, such as Bell states.
Further studies of two coupled Kerr oscillators have led to the
prediction of unconventional photon blockade~\cite{Liew2010},
including its nonreciprocal version~\cite{Li2019}, where even very
weak Kerr nonlinearities can enable high-fidelity generation of
single photons.

Beyond the generation of Fock and Bell states via photon blockade,
Kerr nonlinearity serves as a versatile resource for producing a
broad spectrum of nonclassical states of light~\cite{Tanas2003}.
For example, it enables the creation of highly squeezed states of
light~\cite{Tanas1983, Yamamoto1986, Tanas1991, Bajer2002}, as
well as generation of macroscopically distinguishable quantum
superpositions of coherent states, including the celebrated
Schr\"odinger cat states~\cite{Yurke1986, Tombesi1987} and their
multi-component analogs, often referred to as Schr\"odinger kitten
states~\cite{Miranowicz1990}, which were experimentally generated
in \cite{Kirchmair2013, He2023, Iyama2024}.

The Kerr effect is also fundamental to implementing quantum gates
and performing quantum nondemolition (QND) measurements, where it
facilitates the indirect observation of quantum states without
destroying them~\cite{Milburn1983, Imoto1985, ScullyBook}, as
demonstrated in several landmark experiments
(see~\cite{HarocheBook} for details). Among the various proposals
for Kerr-based quantum gates (see, e.g., \cite{Kanao2022,
Chono2022, Du2024} and references therein), one notable example is
the implementation of fault-tolerant multi-qubit geometric
entangling gates using photonic cat states generated by $N$ Kerr
nonlinear oscillators coupled to a common harmonic resonator
\cite{Chen2022}. That Kerr-based proposal is arguably superior to
other quantum gate implementations based on bosonic codes (see
Table 1 in~\cite{Chen2022}), offering higher gate fidelities and
less demanding coherence requirements in terms of energy
relaxation time ($T_1$) and dephasing time ($T_2$).

Many of these applications critically rely on achieving strong
Kerr nonlinearity at the few-photon level. Several strategies have
been proposed and experimentally explored to enhance this
nonlinearity. In addition to the methods demonstrated in, e.g.,
Refs. \cite{Kirchmair2013, He2023, Iyama2024}, a particularly
noteworthy approach involves sequentially applying two-photon
squeezing processes---governed by the second-order nonlinear
susceptibility $\chi^{(2)}$---to systems with initially weak Kerr
nonlinearity, which arises from the third-order susceptibility
$\chi^{(3)}$~\cite{Bartkowiak2014,Qin2024}. This approach enables,
at least theoretically, an exponential enhancement of the Kerr
nonlinearity.

These capabilities make the Kerr effect indispensable for both
foundational studies in nonlinear optics and the development of
practical quantum technologies.

\section{Model and Its Dynamics}
The dynamics of a system comprising three coupled Kerr oscillators
extends the two-oscillator model previously analyzed in
\cite{Sliwa2012}. Introducing a third oscillator, coupled
nonlinearly and operating at a distinct frequency, enriches the
system's behavior and can be described by the following
Hamiltonian:
\begin{equation}
H = H_0 + H_1 + H_2, \label{eq:hamiltonian}
\end{equation}
where
\begin{equation}
H_0 = \sum_{j=1}^{3} \omega_j a_j^* a_j +
\frac{1}{2}\sum_{j=1}^{3} \epsilon_j (a_j^*)^2 a_j^2,
\label{eq:H0}
\end{equation}
\begin{equation}
H_1 = \epsilon_{12} a_1^* a_2^* a_1 a_2 + \epsilon_{13} a_1^*
a_3^* a_1 a_3 + \epsilon_{23} a_2^* a_3^* a_2 a_3, \label{eq:H1}
\end{equation}
\begin{equation}
H_2 = i\sum_{j=1}^{3} \left[ F_j\left(a_j^* e^{-i\Omega_{jp}t} -
a_j e^{i\Omega_{jp}t}\right) \right]. \label{eq:H2}
\end{equation}
The Hamiltonian $H_0$ describes three independent Kerr
oscillators, where $\omega_j$ denote their natural frequencies and
$\epsilon_j$ quantify the strength of the Kerr nonlinearities. The
Hamiltonian $H_1$ accounts for the nonlinear interactions between
oscillator pairs, with $\epsilon_{12}$, $\epsilon_{13}$, and
$\epsilon_{23}$ representing the respective coupling strengths.
Finally, $H_2$ captures the interaction of each oscillator with
external driving fields, where $F_j$ are the driving amplitudes
and $\Omega_{jp}$ the corresponding pump frequencies.

The equations of motion for the complex variables $a_1$, $a_2$,
and $a_3$ are derived from the Hamiltonian via the relation:
\begin{equation}
   \dot{a}_j = -i\frac{\partial H}{\partial a_j^*} - \gamma_j a_j,
\end{equation}
where the final term accounts for dissipation with $\gamma_j$
denoting the damping rates. Thus, we obtain:
\begin{eqnarray}
\frac{da_1}{dt} &=& -i\omega_1 a_1 - i\epsilon_1 a_1^* a_1^2 -
i\epsilon_{12} a_1 a_2^* a_2
\nonumber \\
&&- i\epsilon_{13} a_1 a_3^* a_3 + F_1 e^{-i\Omega_{1p}t} -
\gamma_1 a_1, \label{eq:motion1}
\end{eqnarray}
\begin{eqnarray}
\frac{da_2}{dt} &=& -i\omega_2 a_2 - i\epsilon_2 a_2^* a_2^2 -
i\epsilon_{12} a_2 a_1^* a_1
\nonumber \\
&&-i\epsilon_{23} a_2 a_3^* a_3 + F_2 e^{-i\Omega_{2p}t} -
\gamma_2 a_2, \label{eq:motion2}
\end{eqnarray}
\begin{eqnarray}
\frac{da_3}{dt} &=& -i\omega_3 a_3 - i\epsilon_3 a_3^* a_3^2 -
i\epsilon_{23} a_3 a_2^* a_2
\nonumber \\
&&- i\epsilon_{13} a_3 a_1^* a_1 + F_3 e^{-i\Omega_{3p}t} -
\gamma_3 a_3. \label{eq:motion3}
\end{eqnarray}
The coupled nonlinear differential equations
(\ref{eq:motion1})--(\ref{eq:motion3}) define a six-dimensional
dynamical system when decomposed into the real and imaginary parts
of each complex variable $a_j$ ($j=1,2,3$), representing the
optical field in each section of the coupler. For each oscillator,
the evolution depends on its intrinsic frequency, Kerr
nonlinearity, nonlinear coupling to the other oscillators,
external driving forces, and energy dissipation.

The damping terms $-\gamma_j a_j$ are essential to our stability
analysis, as they characterize the energy dissipation within the
system. These terms critically influence whether the system's
behavior settles into periodic oscillations or evolves into
chaotic dynamics. For certain parameter regimes, the equations
admit periodic solutions of the form:
\begin{align}
    a_1(t) = \frac{F_1}{\gamma_1} \exp \left[ -i \left( \omega_1 + \epsilon_1 \frac{F_1^2}{\gamma_1^2} \right.\right. &+ \epsilon_{12} \frac{F_2^2}{\gamma_2^2} + \nonumber\\
    & \left.\left. + \epsilon_{13} \frac{F_3^2}{\gamma_3^2} \right) t \right],
    \label{eq:sol1}\\
    a_2(t) = \frac{F_2}{\gamma_2} \exp \left[ -i \left( \omega_2 + \epsilon_2 \frac{F_2^2}{\gamma_2^2} \right.\right. &+ \epsilon_{12} \frac{F_1^2}{\gamma_1^2} + \nonumber\\
    & \left.\left. + \epsilon_{23} \frac{F_3^2}{\gamma_3^2} \right) t \right],
    \label{eq:sol2}\\
    a_3(t) = \frac{F_3}{\gamma_3} \exp \left[ -i \left( \omega_3 + \epsilon_3 \frac{F_3^2}{\gamma_3^2} \right.\right. &+ \epsilon_{23} \frac{F_2^2}{\gamma_2^2} + \nonumber\\
    & \left.\left. + \epsilon_{13} \frac{F_1^2}{\gamma_1^2} \right) t \right],
    \label{eq:sol3}
\end{align}
only if the pumping frequencies are given by:
\begin{eqnarray}
\Omega_{1p} &=& \omega_1 + \epsilon_1 \frac{F_1^2}{\gamma_1^2} + \epsilon_{12} \frac{F_2^2}{\gamma_2^2} + \epsilon_{13} \frac{F_3^2}{\gamma_3^2}, \nonumber \\
\Omega_{2p} &=& \omega_2 + \epsilon_2 \frac{F_2^2}{\gamma_2^2} + \epsilon_{12} \frac{F_1^2}{\gamma_1^2} + \epsilon_{23} \frac{F_3^2}{\gamma_3^2}, \label{eq:eff_freq} \\
\Omega_{3p} &=& \omega_3 + \epsilon_3 \frac{F_3^2}{\gamma_3^2} + \epsilon_{23} \frac{F_2^2}{\gamma_2^2} + \epsilon_{13}
\frac{F_1^2}{\gamma_1^2}. \nonumber
\end{eqnarray}
and the initial conditions are given by:
\begin{equation}
a_{j0} = a_{j}(t=0) = \frac{F_j}{\gamma_j}, 
\quad \text{where} \quad j = 1, 2, 3. 
\label{eq:phase_space0}
\end{equation}
In phase space, these periodic solutions satisfy the following
equations:
\begin{equation}
|a_j|^2 = \frac{F_j^2}{\gamma_j^2}, 
\label{eq:phase_space}
\end{equation}

For our specific analysis, we focus on a system with the following
parameters: $\omega_1 = 1$, $\omega_2 = 0.5$, $\omega_3 = 0.25$,
$\epsilon_1 = \epsilon_2 = \epsilon_3 = 0.01$, $F_1 = F_2 = F_3 =
5$, $\gamma_1 = \gamma_2 = \gamma_3 = 0.5$, $\Omega_{1p} = 2.2$,
$\Omega_{2p} = 1.7$, $\Omega_{3p} = 1.45$, and $\epsilon_{12} =
\epsilon_{13} = \epsilon_{23} = 0.001$ (for the triangular
configuration). With these parameters, the oscillators trace
circular trajectories in phase space, each with a radius of 10 and
oscillating at frequencies of 2.2, 1.7, and 1.45, respectively.

The coupling configurations we examine correspond to different
arrangements of the three oscillators, as illustrated in
Fig.~\ref{fig:fig1}. In the triangular arrangement, each
oscillator is directly coupled to the other two, forming a fully
connected network. In contrast, the sandwich arrangement features
the middle oscillator coupled to both outer oscillators, while the
outer oscillators do not interact directly with each other. These
distinct topologies give rise to markedly different dynamics and
stability characteristics. Notably, these coupling schemes have
direct analogues in quantum computing architectures, where
specific coupling geometries can be engineered to realize targeted
computational functionalities---for example, in the design of
fiber couplers.

Furthermore, the system is six-dimensional with a total of 18
parameters. Exhaustively exploring all parameter combinations is
practically infeasible. Consequently, it is necessary to focus on
a carefully chosen subset of parameters and thoroughly analyze the
system's behavior within that reduced parameter space.

\section{Phase-Space Trajectories and Attractor Structure}
The phase-space analysis of the system
(\ref{eq:motion1})--(\ref{eq:motion3}) reveals key dynamical
properties. For initial conditions $a_{j0}' = \mathrm{Re}\,a_{j0}
= 10$ and $a_{j0}'' = \mathrm{Im}\,a_{j0} = 0$ ($j=1,2,3$), the
phase points of all three subsystems follow circular trajectories
of radius 10, as described by equations
(\ref{eq:sol1})--(\ref{eq:sol3}), with frequencies $\Omega_{1p} =
2.2$, $\Omega_{2p} = 1.7$, and $\Omega_{3p} = 1.45$. However, when
the initial condition of the first subsystem is varied---while
keeping $a_{j0}' = 10$ and $a_{j0}'' = 0$ fixed for
$j=2,3$---multiple attractors emerge.

To systematically analyze this behavior, we performed numerical
simulations varying the initial conditions while keeping the
system parameters fixed. A fourth-order Runge-Kutta method with
adaptive step size control was used to ensure both numerical
stability and accuracy. The integration was carried out up to $t >
1000$ (in normalized units) to allow transient dynamics to decay
and to reliably capture the system's asymptotic behavior.

\begin{figure}[H]
\centering
\includegraphics[width=8.5cm]{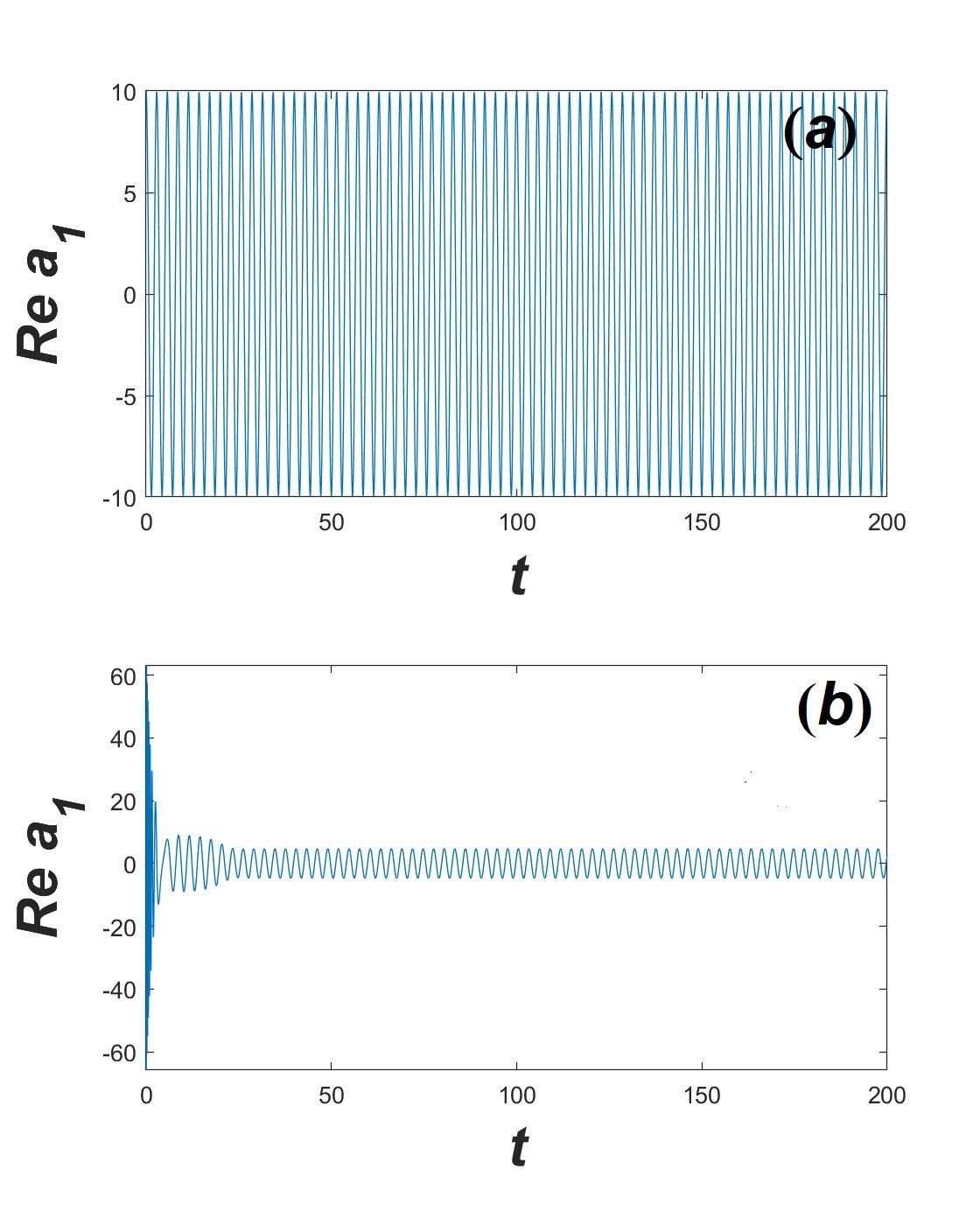}
\caption{Time evolution of $\mathrm{Re}(a_1(t))$ for the
first oscillator subsystem with parameters: $\omega_1 = 1$,
$\omega_2 = 0.5$, $\omega_3 = 0.25$; $\epsilon_1 = \epsilon_2 =
\epsilon_3 = 0.01$; $F_1 = F_2 = F_3 = 5$; $\gamma_1 = \gamma_2 =
\gamma_3 = 0.5$; $\Omega_{1p} = 2.2$, $\Omega_{2p} = 1.7$,
$\Omega_{3p} = 1.45$; and coupling strengths $\epsilon_{12} =
\epsilon_{13} = \epsilon_{23} = 0.001$ (triangular configuration).
Initial conditions are: (a) $\mathrm{Re}\,a_{j0} = 10$ and
$\mathrm{Im}\,a_{j0} = 0$ for $j=1,2,3$; (b) $\mathrm{Re}\,a_{10} = 48$, $\mathrm{Im}\,a_{10} = -48$,
$\mathrm{Re}\,a_{20} = 10$, $\mathrm{Im}\,a_{20} = 0$, $\mathrm{Re}\,a_{30}
= 10$, $\mathrm{Im}\,a_{30} = 0$.} \label{fig:fig2}
\end{figure}

First, we present the time evolution of the first Kerr oscillator---specifically, the real part of $a_1(t)$---for the triangular arrangement. Fig.~\ref{fig:fig2}(a) shows the purely periodic evolution of the oscillator in the case of initial conditions $(\mathrm{Re}\,a_{10}, \mathrm{Im}\,a_{10})$ located on the attractor (i.e., the limit cycle) of radius $10$. When the initial conditions are outside the attractor, transient effects occur before the system reaches the attractor (as in Fig.~\ref{fig:fig2}(b)---in this case, the attractor has a radius of $4.729$).
A similar analysis to that shown in Fig.~\ref{fig:fig2} reveals
that the phase point representing the first subsystem eventually
converges to one of three distinct circular attractors:\\

\noindent
    $\bullet$ the primary attractor with $|a_1|^2 = 100$ (radius $r = 10$),\\
    $\bullet$ the secondary attractor with $|a_1|^2 = (6.457)^2$ (radius $r' = 6.457$), \\
    $\bullet$ the tertiary attractor with $|a_1|^2 =(4.729)^2$ (radius $r'' = 4.729$).
\vspace{0.35cm}

\begin{figure}[H]
\centering
\includegraphics[width=7.2cm]{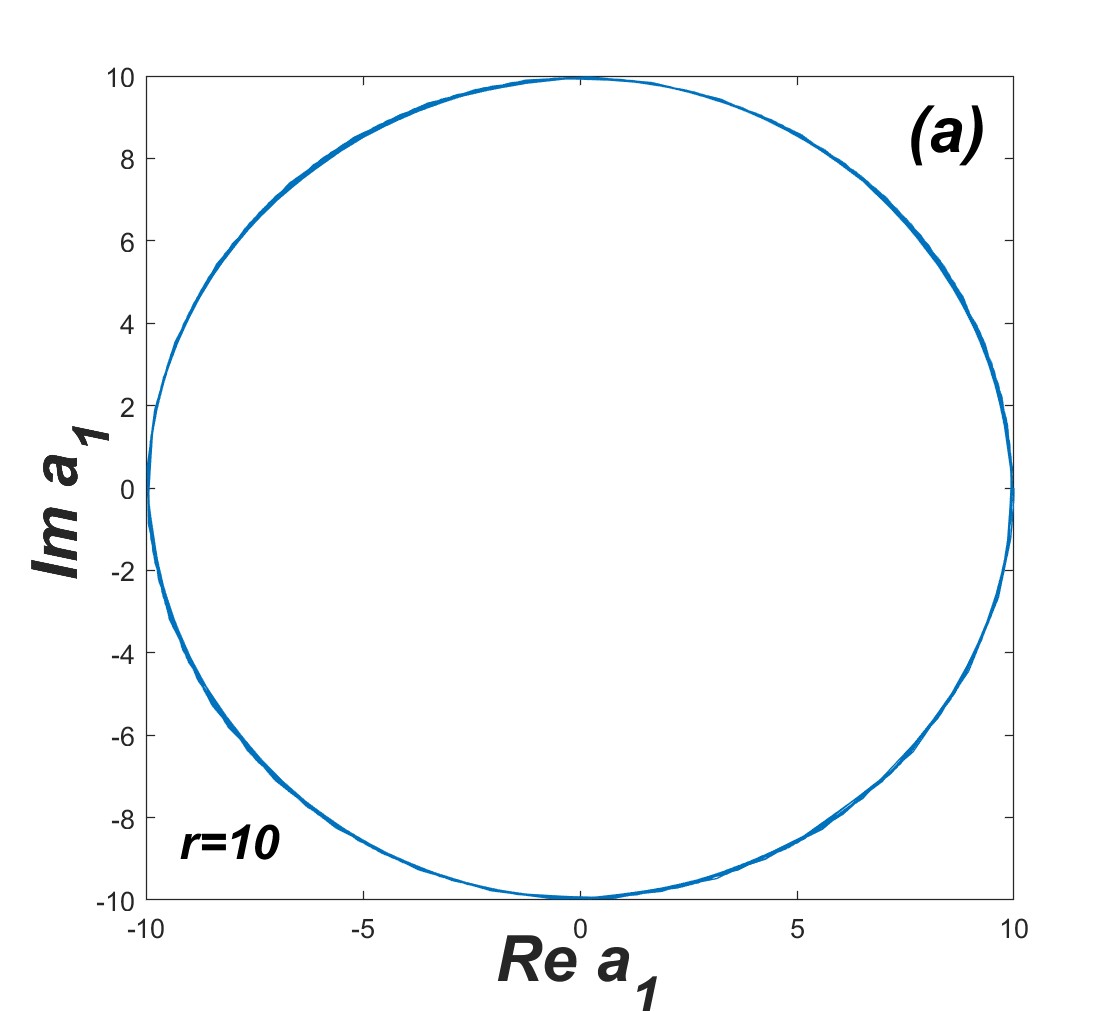}
\includegraphics[width=7.2cm]{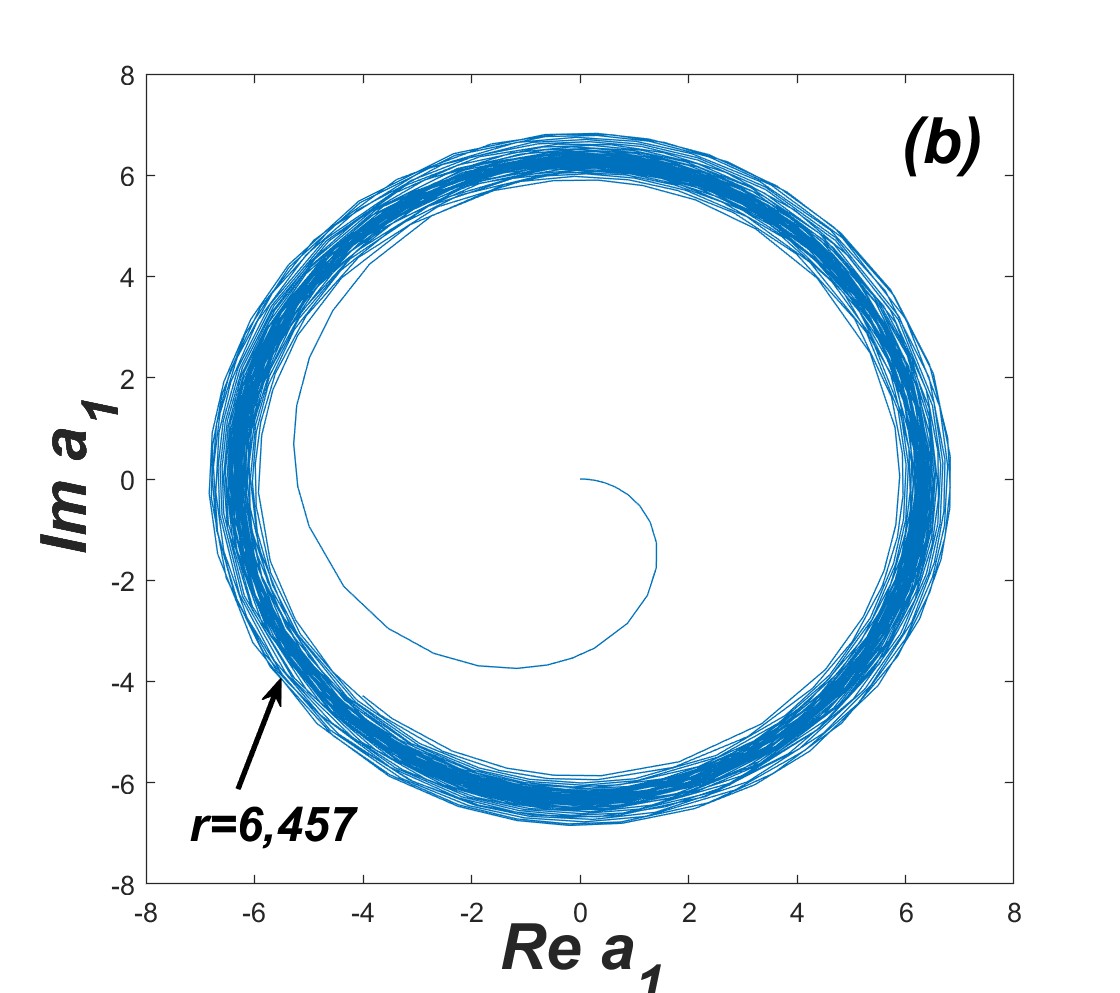}
\includegraphics[width=7.2cm]{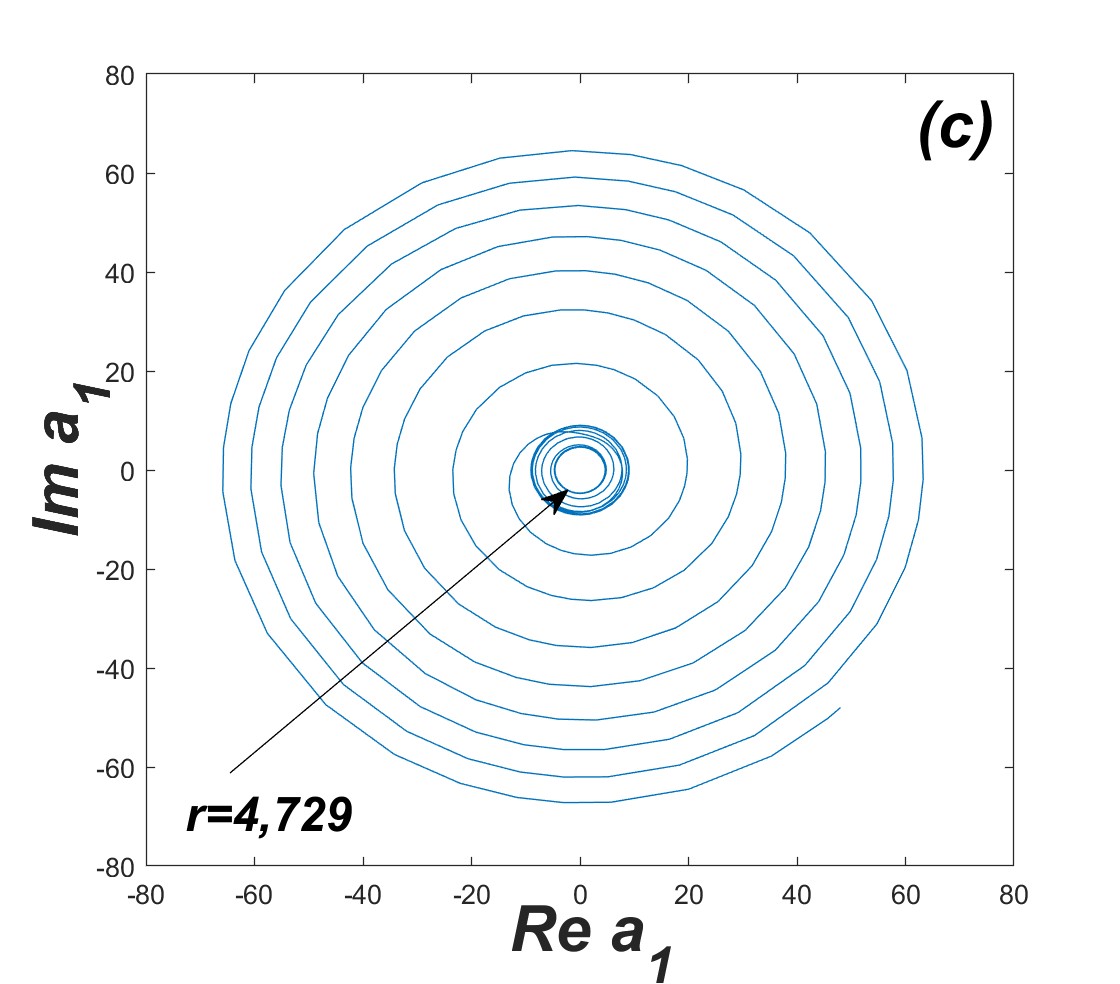}
\caption{Phase-space trajectories of the first Kerr oscillator
with parameters as in Fig.~\ref{fig:fig2}, under the triangular
configuration and varying initial conditions: (a)
$\mathrm{Re}\,a_{10} = 10$, $\mathrm{Im}\,a_{10} = 0$, $\mathrm{Re}\,a_{20}
= 10$, $\mathrm{Im}\,a_{20} = 0$, $\mathrm{Re}\,a_{30} = 10$,
$\mathrm{Im}\,a_{30} = 0$; (b) $\mathrm{Re}\,a_{10} = 0$,
$\mathrm{Im}\,a_{10} = 0$, $\mathrm{Re}\,a_{20} = 10$, $\mathrm{Im}\,a_{20}
= 0$, $\mathrm{Re}\,a_{30} = 10$, $\mathrm{Im}\,a_{30} = 0$; (c)
$\mathrm{Re}\,a_{10} = 48$, $\mathrm{Im}\,a_{10} = -48$,
$\mathrm{Re}\,a_{20} = 10$, $\mathrm{Im}\,a_{20} = 0$, $\mathrm{Re}\,a_{30}
= 10$, $\mathrm{Im}\,a_{30} = 0$. The trajectories demonstrate
convergence toward three distinct attractors with radii: $r = 10$,
$r' = 6.457$, and $r'' = 4.729$, respectively.} \label{fig:fig3}
\end{figure}

Figure~\ref{fig:fig3}(a) shows the phase point starting from
initial conditions $\mathrm{Re}\,a_{10} = 10$, $\mathrm{Im}\,a_{10} =
0$, converging to the primary attractor with radius $r = 10$ by
time $t = 150$. Figure~\ref{fig:fig3}(b) illustrates convergence
to the secondary attractor with radius $r' = 6.457$ from the
initial conditions $\mathrm{Re}\,a_{10} = 0$, $\mathrm{Im}\,a_{10} = 0$.
In Fig.~\ref{fig:fig3}(c), the phase point beginning at
$\mathrm{Re}\,a_{10} = 48$, $\mathrm{Im}\,a_{10} = -48$ converges to the
tertiary attractor with radius $r'' = 4.729$. In the context of
quantum computing, these distinct stable states can correspond to
different computational states in multi-state quantum memory.

\begin{figure}[H]
\centering
\includegraphics[width=\columnwidth]{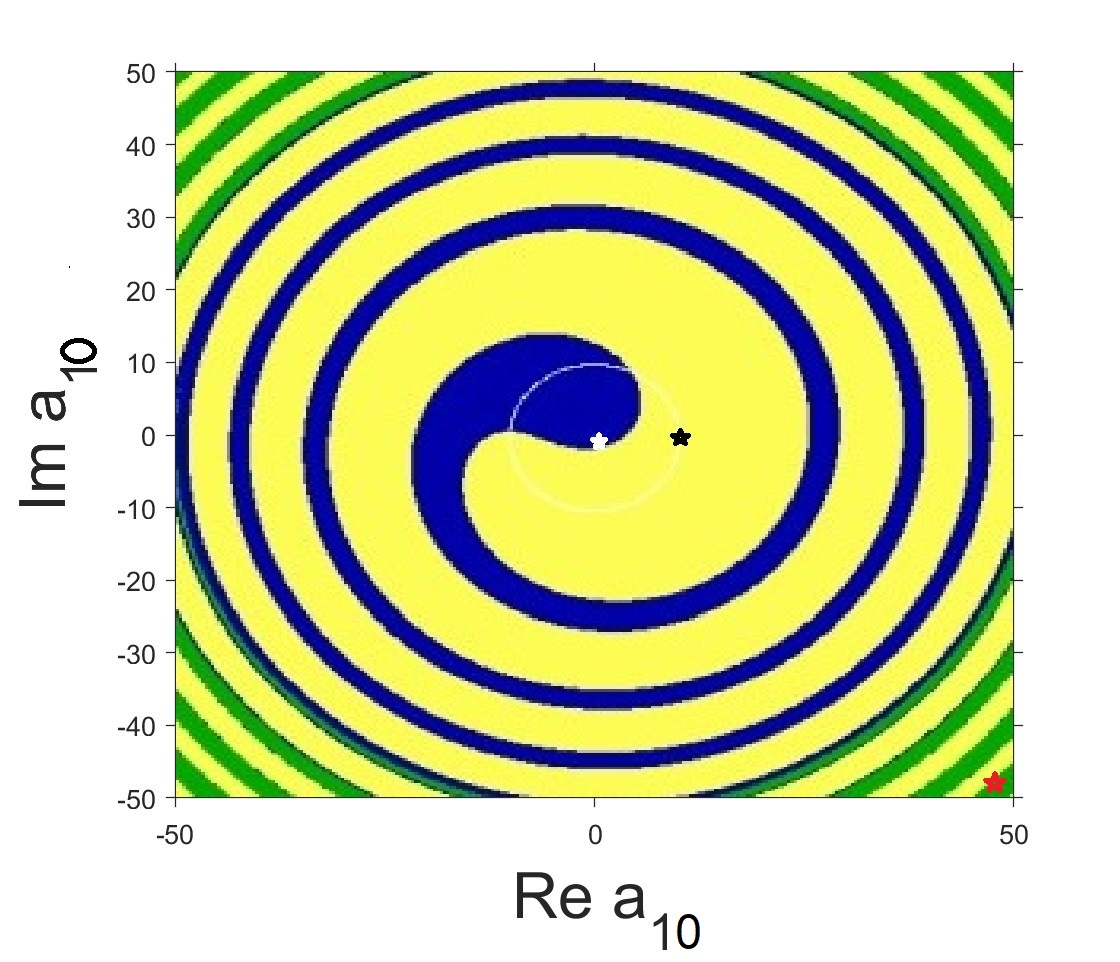}
\caption{Basin of attraction corresponding to the case shown in
Fig.~\ref{fig:fig3}. Colors indicate which stable attractor the
subsystem consisting of the first oscillator reaches from each
initial condition $(\mathrm{Re}\,a_{10}, \mathrm{Im}\,a_{10})$: $r = 10$ (yellow), $r
= 6.457$ (blue), and $r = 4.729$ (green). Asterisks mark the exact
initial conditions used in Fig.~\ref{fig:fig3}(a--c). Note that
only the attractor with radius $r = 10$ is explicitly labeled in
the figure.} \label{fig:fig-basen2}
\end{figure}

The basins of attraction for each attractor were mapped by
sampling a grid of initial conditions in the $(\mathrm{Re}\,a_{10},
\mathrm{Im}\,a_{10})$ plane, as shown in Fig.~\ref{fig:fig-basen2}. We
observed that the basin boundaries exhibit fractal-like
structures, reflecting a high sensitivity to initial
conditions---a hallmark of nonlinear systems with multiple
attractors.
\begin{table}[H]
\centering
\begin{tabular}{lcccccc}
Configuration & $r$ & $r'$ & $r''$ & $\Omega_{1p}$ & $\Omega_{2p}$ & $\Omega_{3p}$ \\
\hline triangular
& $10$ & $6.457$ & $4.729$ & $2.2$ & $1.7$ & $1.45$ \\
sandwich
& $10$ & $6.73$ & $5.339$ & $2.1$ & $1.7$ & $1.35$ \\
\end{tabular}
\caption{\label{tab:attractors} Attractor radii and frequencies
for different coupling configurations, specifically the triangular
arrangement ($\epsilon_{12} = \epsilon_{13} = \epsilon_{23} =
0.001$) and the sandwich arrangement ($\epsilon_{13} = 0$,
$\epsilon_{12} = \epsilon_{23} = 0.001$). }
\end{table}
To investigate the influence of coupling configuration, we
compared the triangular arrangement with the sandwich
configuration, in which $\epsilon_{13} = 0$, meaning there is no
direct coupling between the first and third oscillators.
Table~\ref{tab:attractors} summarizes the properties of the
attractors for each configuration.

While both configurations exhibit similar transient behavior and
ultimately converge to circular periodic orbits, a notable
difference emerges: the sandwich configuration results in larger
secondary and tertiary attractors than the triangular one. This
counterintuitive finding suggests that reducing the number of
couplings can, in certain cases, enhance rather than suppress the
intensity of specific oscillation modes---highlighting the
intricate and complex nature of nonlinear interactions in the
system. It is also important to note that reducing the number of
couplings introduces greater asymmetry into the system, which may
play a significant role in shaping the dynamical properties of the
coupler. This finding has significant implications for designing
quantum computing architectures where specific coupling geometries
can be engineered to achieve desired computational properties.

\section{Lyapunov Exponent Analysis and Transition to Chaos}

Lyapunov exponents offer a rigorous means of characterizing the
stability of a dynamical system by quantifying the rate at which
nearby trajectories in phase space diverge or converge. A positive
maximal Lyapunov exponent signifies exponential divergence of
initially close trajectories---a hallmark of chaotic behavior. In
contrast, a zero maximal exponent indicates quasiperiodic
dynamics, while a negative maximal exponent corresponds to
periodic behavior, where trajectories remain bounded and converge.
The method applied ranks the Lyapunov exponents in descending
order. If the largest exponent is positive, the system is chaotic.
If two or more exponents are positive, the system is classified as
hyperchaotic, exhibiting even more complex instability.

In quantum computing applications, understanding the stability
characteristics of Kerr oscillator systems is crucial for
designing reliable quantum operations, as chaos can lead to
information loss and decoherence.

For our system of three nonlinearly coupled Kerr oscillators, we
employed the method of Wolf et al. \cite{Wolf1985}, which
incorporates the Gram-Schmidt reorthonormalization (GSR)
algorithm. This method tracks the evolution of perturbation
vectors in the tangent space alongside the phase-space trajectory,
periodically reorthonormalizing the basis vectors to avoid
numerical instability and ensure accurate computation of Lyapunov
exponents.

The system of equations (\ref{eq:motion1})--(\ref{eq:motion3})
defines a six-dimensional dynamical system. Linearizing around a
reference trajectory yields an additional system of 36 variational
equations---corresponding to 6 perturbation vectors in
6-dimensional space---required to compute the full Lyapunov
spectrum. Consequently, analyzing the three-oscillator Kerr
coupler involves solving a total of 42 coupled ordinary
differential equations (ODEs).

Our numerical implementation involved the following steps:
\begin{enumerate}
\item
Simultaneous integration of the original system defined by
Eqs.~(\ref{eq:motion1})--(\ref{eq:motion3}) along with the
corresponding set of linearized variational equations.
\item
Periodic application of the GSR procedure, typically every 0.01
time units, to maintain numerical stability of the tangent space
vectors.
\item
Accumulation of logarithmic rates of expansion and contraction
along each orthonormal direction in phase space.
\item
Long-time averaging of the accumulated rates, typically over more
than 5000 time units, to ensure convergence and to eliminate
transient effects.
\end{enumerate}
To ensure the robustness of our numerical procedure, we validated
the results by confirming that the sum of all Lyapunov exponents
approximates the theoretical expectation of $-(\gamma_1 +
\gamma_2 + \gamma_3)$, which reflects the total dissipation in the
system.

The identification of distinct stability regimes is particularly
relevant for quantum computing applications, where controlled
chaotic behavior can be exploited for specific tasks such as
random number generation and reservoir computing.

Given the high dimensionality and large parameter space of the
system, all Lyapunov spectrum calculations were performed using a
fixed set of baseline parameters: $\omega_1 = 1$, $\omega_2 =
0.5$, $\omega_3 = 0.25$, $\epsilon_1 = \epsilon_2 = \epsilon_3 = 0.01$, $F_1 = F_2 = F_3 = 5$, $\Omega_{2p} = 1.7$, and $\Omega_{3p} = 1.45$. The initial conditions were as follows: $\mathrm{Re}\,a_{j0} = 10$, $\mathrm{Im}\,a_{j0} = 0$ ($j = 1, 2, 3$).
\begin{figure*}[t!]
\centering
\includegraphics[width=8cm]{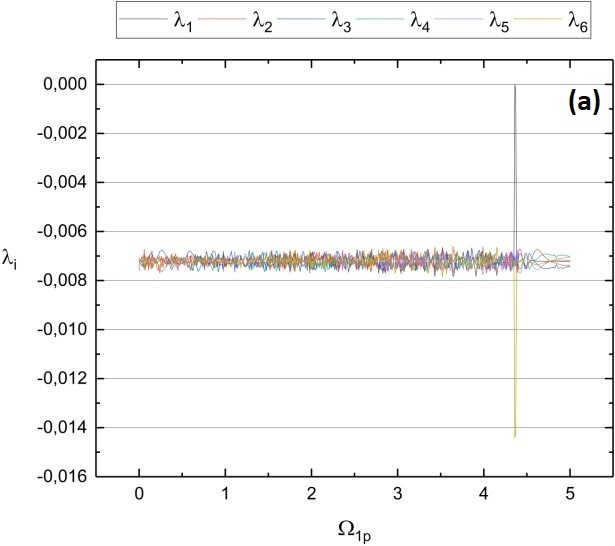}
\includegraphics[width=8cm]{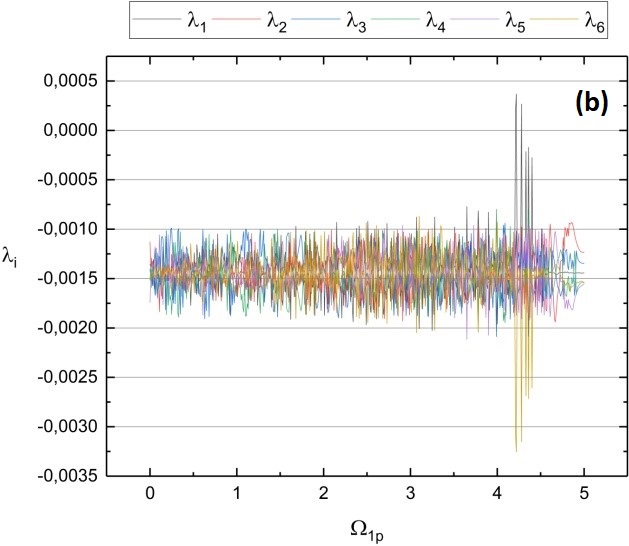}
\includegraphics[width=8cm]{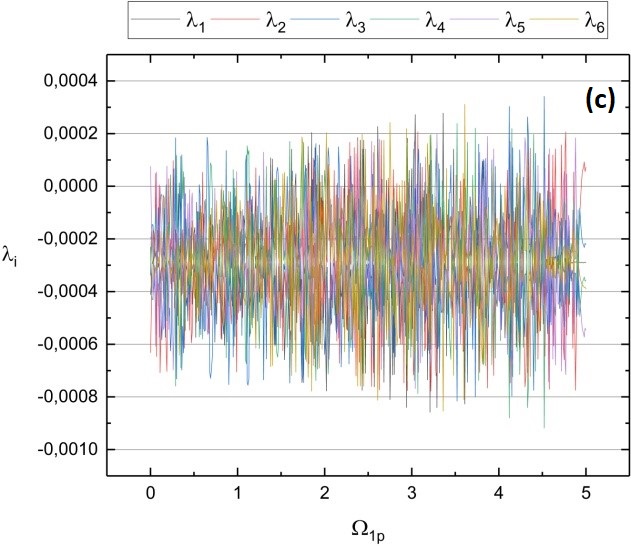}
\includegraphics[width=8cm]{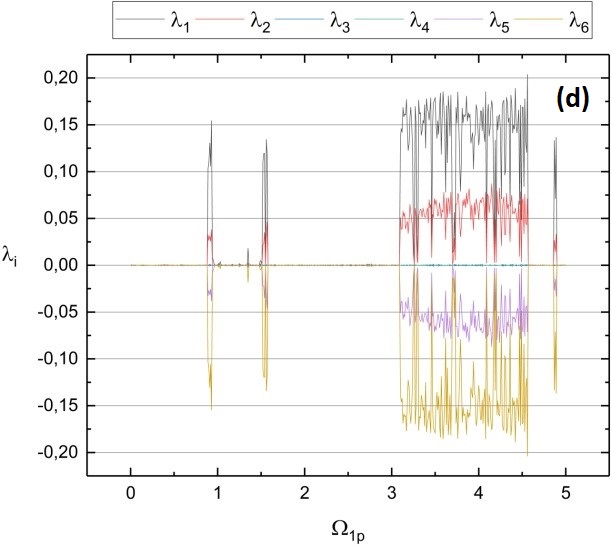}
\caption{Lyapunov exponents $\lambda_1$--$\lambda_6$ for the
triangular arrangement ($\epsilon_{12} =
\epsilon_{13} = \epsilon_{23} = 0.001$) of the Kerr couplers as a function of pumping
frequency $\Omega_{1p}$ of the first oscillator and for different
damping constants: (a) $\gamma_1 = \gamma_2 = \gamma_3 = \gamma =
0.005$, (b) $\gamma = 0.001$, (c) $\gamma = 0.0002$, (d) $\gamma =
0$. Other system parameters are the same as in
Fig.~\ref{fig:fig2}. Note the qualitative change in system
behavior as the damping decreases: in panel (d), the emergence of
positive Lyapunov exponents signals the onset of chaotic---and
potentially hyperchaotic---dynamics.} \label{fig:fig4}
\end{figure*}
Figure~\ref{fig:fig4} presents the key results of our analysis of
the Lyapunov exponent spectrum for the triangular configuration ($\epsilon_{12} = \epsilon_{13} = \epsilon_{23} = 0.001$),
shown as a function of the pumping frequency $\Omega_{1p}$ of the
first Kerr oscillator. Each panel corresponds to a different value
of the damping constants, with $\gamma_1 = \gamma_2 = \gamma_3 =
\gamma$. Several critical observations emerge from this analysis:
\begin{enumerate}
\item \textbf{Strong damping regime} (Fig.~\ref{fig:fig4}(a) for $\gamma = 0.005$)
\noindent All Lyapunov exponents remain negative and approximately
constant across the entire frequency range, indicating a strongly
dissipative regime. In this regime, the system consistently
converges to stable fixed points or limit cycles, regardless of
the pumping frequency. From a quantum computing perspective, such
behavior corresponds to robust and predictable dynamics---an
essential feature for implementing high-fidelity quantum gates.

\item \textbf{Intermediate damping regime} (Fig.~\ref{fig:fig4}(b) for $\gamma=0.001$)
\noindent Although all Lyapunov exponents remain negative and show similar variation across the pumping frequency, there are more points of rapidly increasing exponents, indicating
elevated sensitivity to parameter changes. The smaller magnitudes
of the exponents imply slower convergence to attractors. This
regime may be advantageous for applications that benefit from
heightened sensitivity to inputs, such as quantum sensing.

\item \textbf{Weak damping regime} (Fig.~\ref{fig:fig4}(c) for $\gamma=0.0002$) \\
\noindent The largest Lyapunov exponents approach zero at specific
frequencies, indicating that the system is near critical
transitions. The spectrum exhibits pronounced frequency
dependence, with fluctuations reflecting competing dynamical
regimes. Detailed analysis reveals a frequent occurrence of
quasi-periodic states. Similar ``edge-of-chaos'' regimes have been
leveraged in recent quantum neural network implementations to
enhance computational capacity.

\item \textbf{Undamped regime} (Fig.~\ref{fig:fig4}(d) for $\gamma = 0$) \\
\noindent A critical transition occurs in this regime, where some
Lyapunov exponents become positive within specific frequency
intervals, confirming the onset of chaotic behavior. These chaotic
regimes are interspersed with regular (non-chaotic) windows,
exhibiting intermittency---a hallmark of many nonlinear systems
that can be harnessed for chaos-based computing applications.
Moreover, the presence of extensive frequency ranges in
$\Omega_{1p}$ supporting hyperchaotic dynamics highlights the
system's potential usefulness for cryptographic applications and
secure communications.
\end{enumerate}

The transition to chaos as damping decreases can be understood as
a competition between energy dissipation and nonlinear energy
transfer combined with external pumping. When damping is
sufficiently strong, energy dissipation dominates, suppressing
nonlinear mode interactions and maintaining stable system
behavior. As damping weakens, nonlinear coupling and pumping
effects gain prominence, facilitating nontrivial energy exchanges
that can lead to chaotic dynamics once dissipation is no longer
able to contain them. In the context of quantum hardware, this
insight is valuable for designing dissipation engineering
strategies aimed either at preserving system stability or at
deliberately inducing controlled chaos for specialized
applications.

\begin{figure*}[t!]
\centering
\includegraphics[width=8cm]{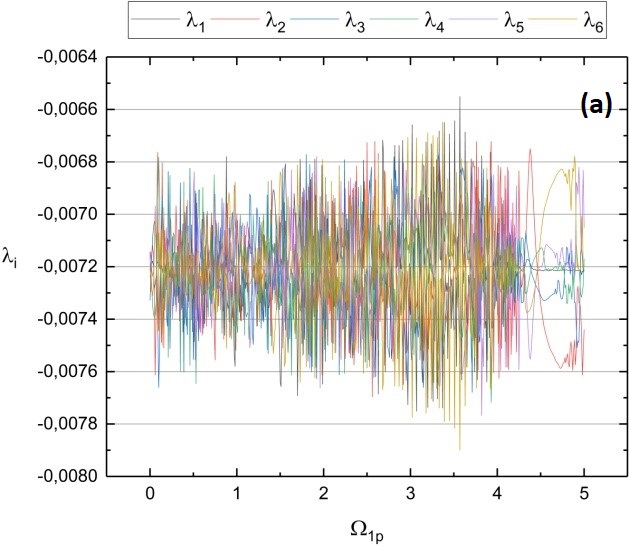}
\includegraphics[width=8cm]{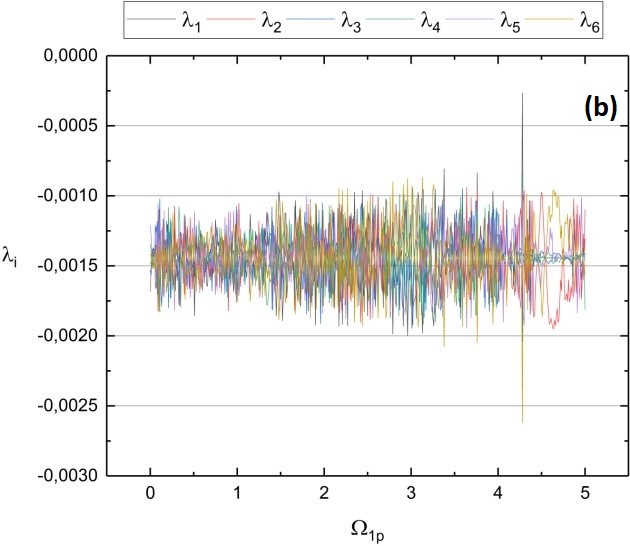}
\includegraphics[width=8cm]{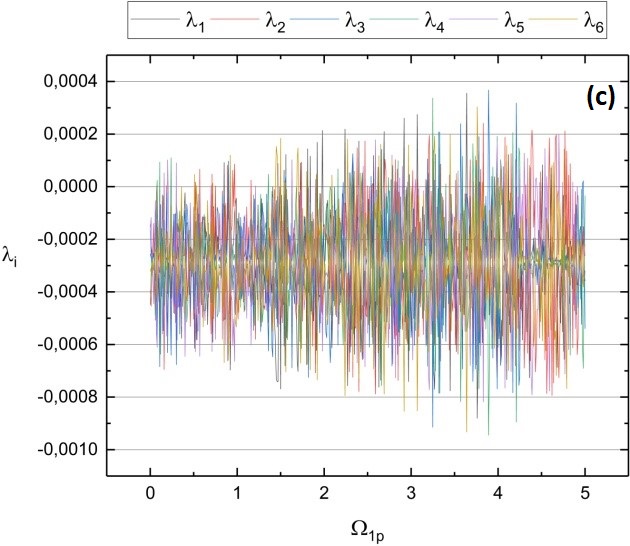}
\includegraphics[width=8cm]{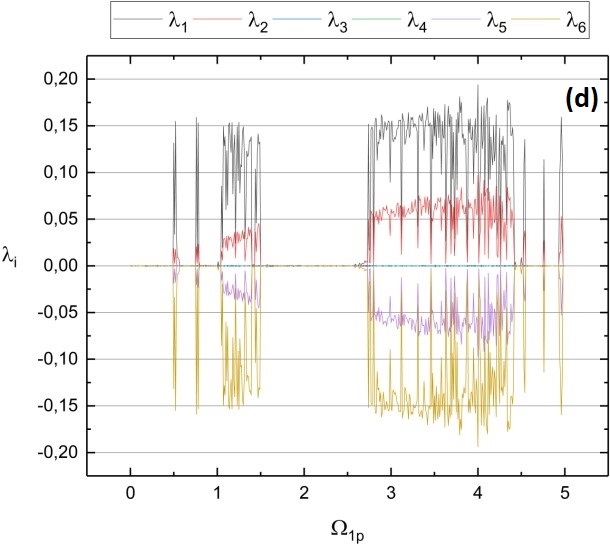}
\caption{Lyapunov exponents $\lambda_1$--$\lambda_6$ for the
sandwich arrangement ($\epsilon_{13}=0, \epsilon_{12} =
\epsilon_{23} = 0.001$) of the Kerr couplers as a function of
pumping frequency $\Omega_{1p}$ of the first oscillator and for
different damping constants: (a) $\gamma_1 = \gamma_2 = \gamma_3 =
\gamma = 0.005$, (b) $\gamma = 0.001$, (c) $\gamma = 0.0002$, (d)
$\gamma = 0$. Other system parameters are the same as in
Fig.~\ref{fig:fig2}.} \label{fig:fig5}
\end{figure*}

A similar stability analysis was carried out for the sandwich
configuration shown in Fig.~\ref{fig:fig1} (with $\epsilon_{13} =
0$). Under parameters analogous to those in Fig.~\ref{fig:fig4},
the system exhibits comparable behavior, as illustrated in
Fig.~\ref{fig:fig5}. Notably, the regions of chaos and hyperchaos
expand in the absence of damping, reflecting a reduction in
overall stability. This decreased stability arises from the
increased asymmetry of the system: in the sandwich arrangement,
the central Kerr oscillator couples to both neighbors, whereas
each outer oscillator is coupled to only one neighbor.
\begin{figure*}[t!]
\centering
\includegraphics[width=8cm]{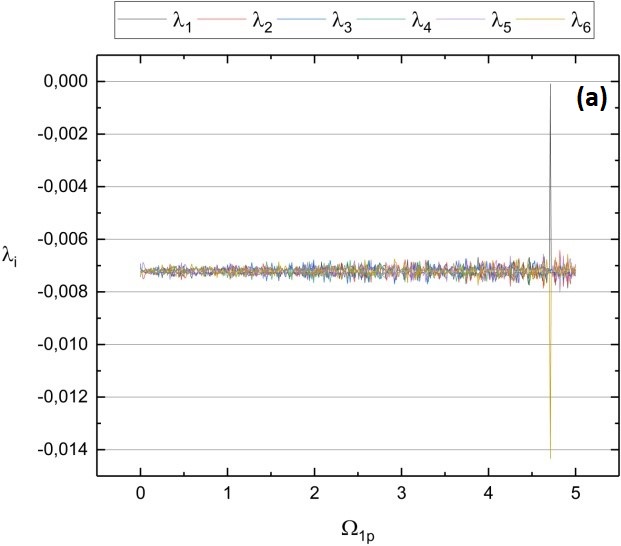}
\includegraphics[width=8cm]{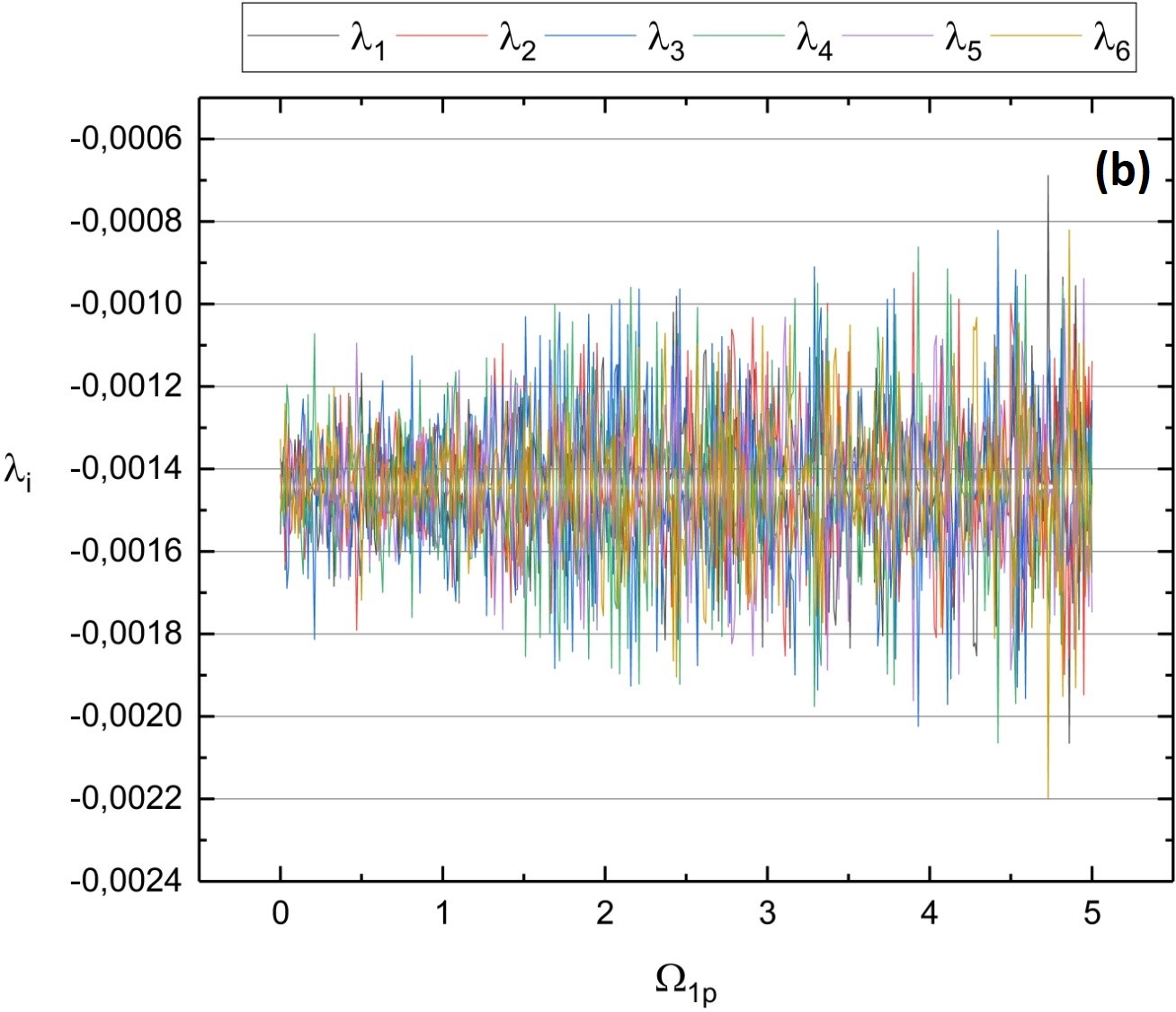}
\includegraphics[width=8cm]{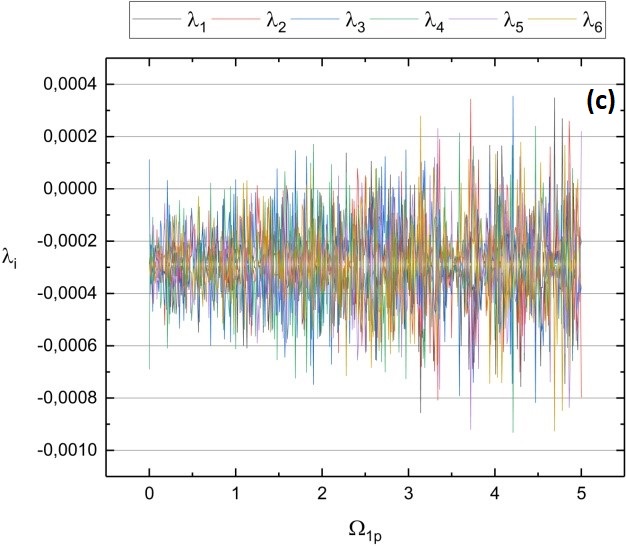}
\includegraphics[width=8cm]{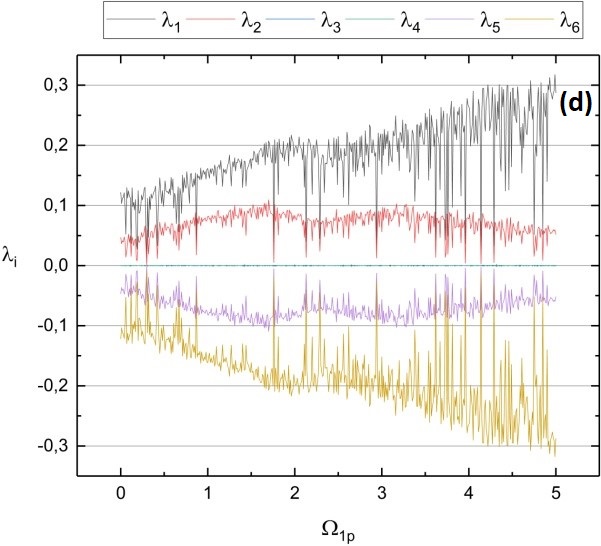}
\caption{Lyapunov exponents $\lambda_1$--$\lambda_6$ for the
sandwich arrangement of the Kerr couplers under strong coupling
conditions ($\epsilon_{13} = 0, \epsilon_{12} = \epsilon_{23} = 0.01$)
as a function of
pumping frequency $\Omega_{1p}$ of the first oscillator and for
different damping constants: (a) $\gamma_1 = \gamma_2 = \gamma_3 =
\gamma = 0.005$, (b) $\gamma = 0.001$, (c) $\gamma = 0.0002$, (d)
$\gamma = 0$. Other system parameters are the same as in
Fig.~\ref{fig:fig2}.} \label{fig:fig6}
\end{figure*}

To confirm that the observed chaotic behavior is genuine and not a
numerical artifact, we performed several validation tests:
\begin{itemize}
\item
Varying the integration step size and GSR intervals to verify
numerical convergence.

\item
Testing multiple sets of initial conditions to ensure consistent
Lyapunov spectra across simulations.

\item Calculating the correlation dimension, which confirmed the fractal
nature of the attractors in the chaotic regimes.
\end{itemize}

Furthermore, we investigated how the transition to chaos depends
on coupling strengths. We found that increasing the coupling
parameters $\epsilon_{12}$, $\epsilon_{13}$, and $\epsilon_{23}$
lowers the critical damping threshold for chaos, confirming that
the nonlinear coupling is indeed the mechanism driving the chaotic
behavior. This is particularly visible for the sandwich
configuration, which confirms the leading role of not only the
coupling but also the asymmetry of the coupler system
(Fig.~\ref{fig:fig6}). The Lyapunov exponents clearly indicate a
marked increase in the system's instability
(Fig.~\ref{fig:fig6}(d)). This finding is particularly relevant for
quantum computing implementations, where coupling strengths can be
precisely controlled to achieve desired stability characteristics.

\section{Chaotic Beats}
In certain coupled nonlinear systems, it is possible to observe a
distinctive dynamical behavior known as chaotic beats. This
phenomenon was first numerically identified in a system of two
coupled Kerr and Duffing oscillators \cite{grygiel2002}. Since
then, chaotic beats have been reported in various systems,
including Chua's circuit \cite{cofagna2004}, coupled
second-harmonic generators of light \cite{sliwa2007}, and
memristive-driven Chua circuits \cite{Ahamed2011}. Notably, the
phenomenon has also been demonstrated experimentally in an
electronic setup consisting of two forced dissipative LCR
oscillators sharing a nonlinear element \cite{Asird2016}.

\begin{figure*}[t!]
\centering
\includegraphics[width=8cm]{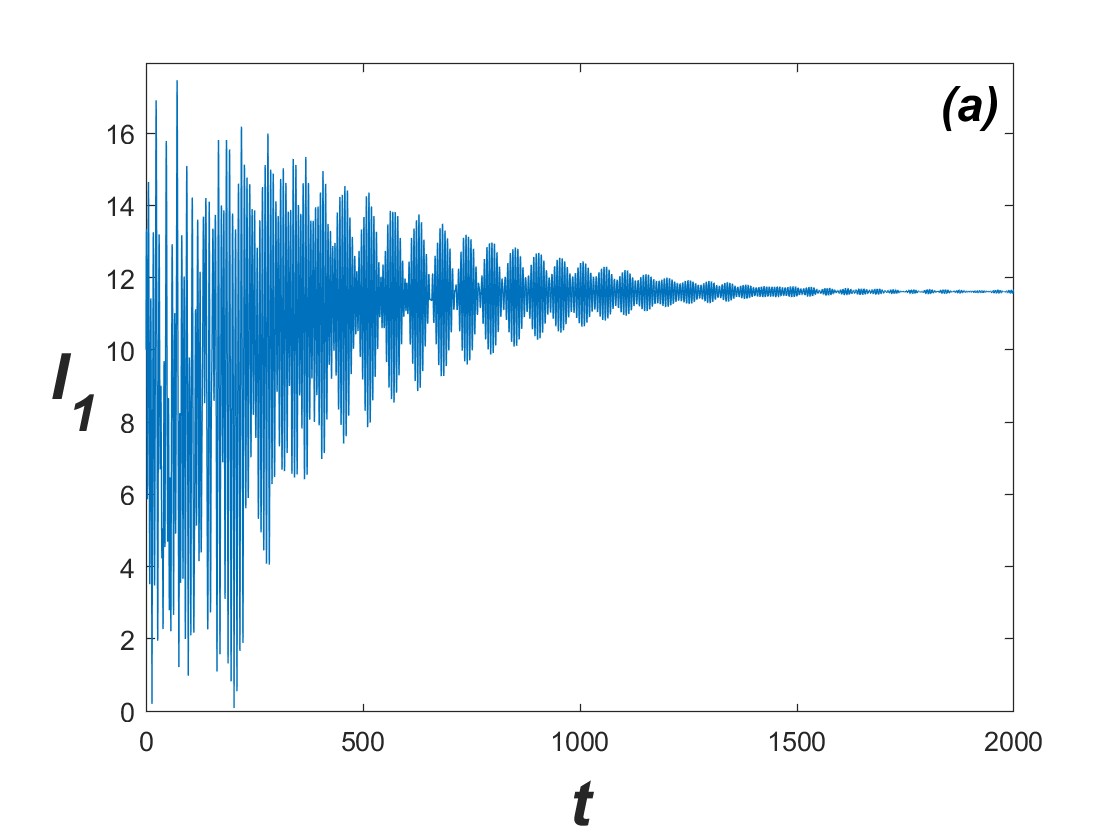}
\includegraphics[width=8cm]{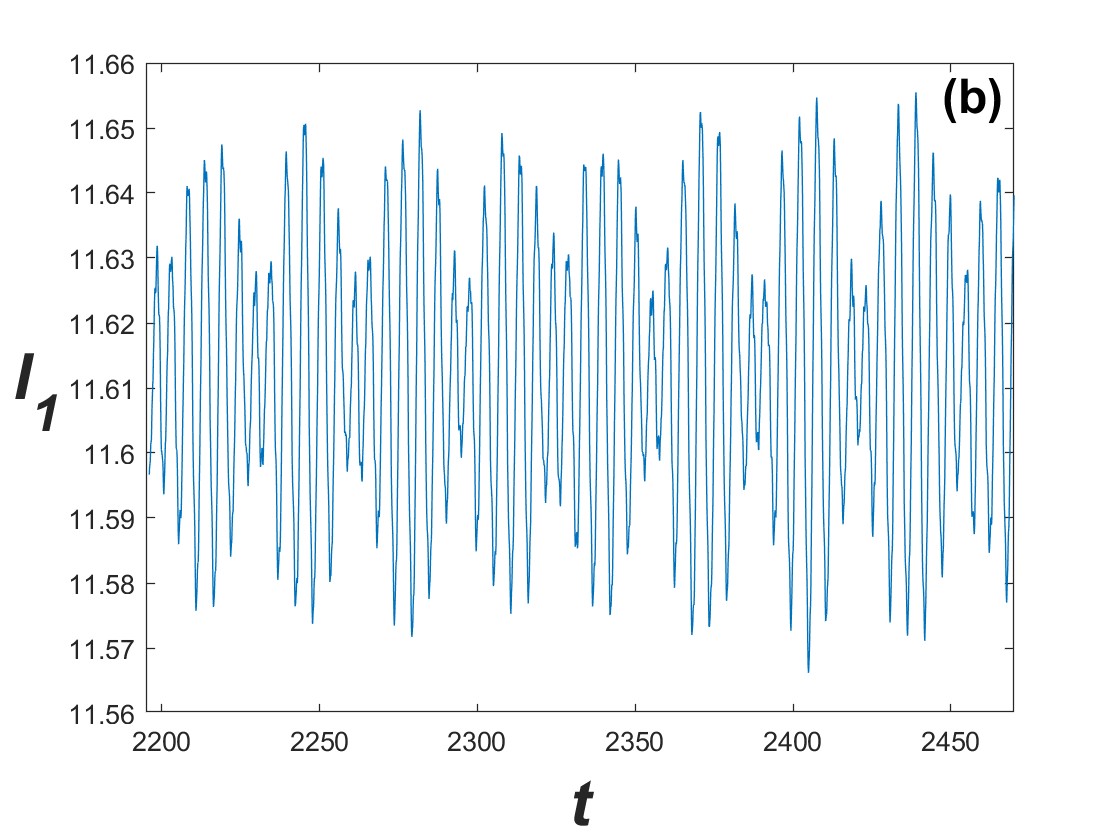}
\caption{Chaotic beats in the first Kerr oscillator. Time
evolution of the intensity $I_1(t) = |a_1(t)|^2$ is shown for the
case of strong coupling in the sandwich configuration, with
pumping frequency $\Omega_{1p} = 1$. All other parameters are the
same as in Fig.~\ref{fig:fig6}(d). Panel (a) illustrates the
initial strongly chaotic transient regime, while panel (b)
displays the subsequent emergence of a stationary chaotic beat
pattern.} \label{fig:fig7}
\end{figure*}

In general, chaotic beats refer to signals where the envelope of
amplitude modulation exhibits chaotic fluctuations, while the
underlying carrier frequency remains nearly constant. This
phenomenon typically arises in weakly coupled nonlinear systems.
Interestingly, in the case of a system of three coupled Kerr
oscillators, we identified a specific set of parameters for which
chaotic beats emerge even under strong coupling conditions. In
this configuration, the intensity of the first oscillator, defined
as $I_1(t) = |a_1(t)|^2$, evolves---after an initial period of
strongly chaotic transients---into a regime of persistent,
stationary-like chaotic beats, as illustrated in
Fig.~\ref{fig:fig7}.
Further analysis reveals that, as the value of the pumping
frequency $\Omega_{1p}$ increases, the system gradually loses its
beat-like characteristics and transitions into a regime of purely
chaotic behavior.

\section{Quantum Computing Applications and Experimental Implementations}
Our analysis of three coupled Kerr oscillators carries important
implications for emerging quantum computing technologies. Although
the present treatment is classical, the stability regimes we have
identified remain highly relevant for mesoscopic and macroscopic
systems where classical and quantum types of behavior coexist. In
this section, we explore the connection between our results and
recent advancements in quantum computing, particularly in the
context of experimental platforms that utilize nonlinear
oscillators and engineered dissipation.

\subsection{Advantages of three-oscillator systems}

Three coupled Kerr oscillators represent a critical minimum
configuration for several quantum computing applications that
cannot be realized with simpler two-oscillator systems:
\begin{enumerate}
\item \textbf{Enhanced computational basis:}
The multiple stable attractors we identified (as summarized in
Tab.~\ref{tab:attractors}) provide an expanded state space for
information encoding. In quantum computing implementations based
on Kerr parametric oscillators (KPOs), these states can represent
distinct computational basis states \cite{Margiani2025}.

\item \textbf{Triangular coupling topology:}
The triangular configuration enables genuine three-body
interactions that cannot arise in systems with only two
oscillators. Recent work by Margiani et al. \cite{Margiani2025}
demonstrated that a system of three strongly coupled KPOs can
function as a Boltzmann machine capable of simulating Ising
Hamiltonians. This architecture has direct applications in solving
combinatorial optimization problems and highlights the
computational potential of nonlinear oscillator networks.

\item \textbf{Error correction capabilities:}
Systems composed of three coupled oscillators support redundant
encoding schemes that enhance robustness against noise and
decoherence---key requirements for scalable quantum computing. Our
stability analysis identifies the parameter regimes in which such
error-resilient encoding is most effective, offering guidance for
the design of quantum architectures with improved fault tolerance.
\end{enumerate}

\subsection{Experimental platforms}
Recent experimental advances have made the implementation of
coupled Kerr oscillator systems increasingly feasible:
\begin{enumerate}
\item \textbf{Superconducting circuits:}
Superconducting circuits have emerged as a leading platform for
realizing coupled Kerr parametric oscillators (KPOs) in quantum
computing \cite{Gu2017, Yin2021}. Recent experiments have
demonstrated high-fidelity quantum gate operations using KPOs,
including $R_x$ gates via parity-selective transitions
\cite{Kanao2022}, and two-qubit $R_{zz}$ gates with fidelities
exceeding 99.9\% in systems of highly detuned KPOs
\cite{Chono2022}.

\item \textbf{Integrated photonics:}
Silicon nitride microresonators have demonstrated high-efficiency
optical parametric oscillation with conversion efficiencies
reaching 29\% \cite{Perez2022}. These platforms benefit from
scalability and compatibility with existing semiconductor
manufacturing technologies. For example, in situ control of
integrated Kerr nonlinearity with a tuning range of 10~dB has
recently been demonstrated \cite{Cui2022}, enabling dynamic
modulation of nonlinear interactions in superconducting quantum
circuits.

\item \textbf{Commercial implementations:}
IBM's Quantum System Two, introduced in 2023, marks a significant
milestone in the commercial advancement of quantum processors
based on coupled nonlinear oscillators. The system is capable of
executing up to 1800 quantum gates within coherence
times---nearly quadrupling the capacity of previous-generation
devices \cite{IBM2023}.
\end{enumerate}

The critical damping thresholds identified in Section V offer
valuable guidance for experimental implementations by delineating
parameter regimes that ensure stable operation versus those prone
to chaotic transitions. This insight is especially pertinent for
superconducting circuit platforms, where damping rates can be
precisely engineered.

\subsection{Potential applications for quantum computing}

The distinct dynamical regimes revealed by our Lyapunov exponent
analysis correspond to specific operational modes with direct
applications in quantum computing:
\begin{enumerate}
\item \textbf{Quantum gates:}
The stable regime characterized by negative Lyapunov exponents is
ideal for implementing reliable quantum gates. Recent experiments
have demonstrated that KPOs can perform both high-fidelity
single-qubit operations and entangling gates \cite{Kanao2022}.

\item \textbf{Quantum neural networks:}
The near-critical regime, where Lyapunov exponents approach zero
yet remain negative (Fig.~\ref{fig:fig4}(c)), offers enhanced
computational capacity well-suited for quantum neural networks.
Recent studies have demonstrated that even with just two coupled
quantum oscillators, a quantum reservoir containing up to 81
effective neurons can be realized, achieving 99\% accuracy on
benchmark tasks \cite{Brunner2013}.

\item \textbf{Chaos-based computing:}
The chaotic regime with positive Lyapunov exponents can be
exploited for specialized computing tasks, including quantum
random number generation and quantum cryptography. Controlled
chaotic behavior in optical systems has been demonstrated as an
effective mechanism for generating high-entropy random bit streams
\cite{Uchida2008}.
\end{enumerate}

Our analysis of the impact of coupling configurations on attractor
properties (Table~\ref{tab:attractors}) is particularly relevant
for quantum computing applications that demand precise control
over system dynamics. Notably, the observation that sandwich
configurations support larger secondary and tertiary attractors
indicates that the deliberate removal of specific couplings can
enhance particular computational functionalities.

\section{Comparative analysis and future work}

\subsection{Comparison with two-oscillator systems}

Our three-oscillator system shares certain features with the
two-oscillator case studied by \'Sliwa and Grygiel
\cite{Sliwa2012}, such as multiple attractors and
parameter-dependent dynamics. However, the addition of a third
oscillator gives rise to novel phenomena and richer dynamical
behavior, including:
\begin{enumerate}
\item \textbf{Increased attractor complexity:}
The three-oscillator system supports a richer set of attractors,
including the tertiary attractor not observed in the
two-oscillator case. This can be attributed to the additional
degrees of freedom and coupling pathways.

\item \textbf{Configuration-dependent dynamics:}
The triangular versus sandwich configurations exhibit distinct
dynamical properties, with sandwich configurations supporting
larger secondary and tertiary attractors. This indicates that in
certain parameter regimes, reduced coupling can
counter-intuitively enhance intensity of the process.

\item \textbf{Lower chaos threshold:}
Compared to the two-oscillator system, our three-oscillator system
transitions to chaos at higher damping values, indicating
increased dynamical complexity.
\end{enumerate}

\subsection{Physical mechanisms}

The multiple attractors observed in our system arise from
nonlinear mode competition. The nonlinear coupling terms in Eqs.
(\ref{eq:motion1})--(\ref{eq:motion3}) facilitate energy exchange
between oscillators, creating a complex energy landscape with
multiple local minima that correspond to distinct stable
oscillation patterns.

The transition to chaos as damping decreases reflects the delicate
balance between energy dissipation and nonlinear energy transfer.
When damping is sufficiently strong, dissipation dominates,
yielding simple and stable attractor structures. As damping
weakens, nonlinear energy transfer gains prominence, ultimately
driving the system into chaotic dynamics once dissipation can no
longer offset these nonlinear effects.

\subsection{Connection to quantum-classical correspondence}

While our analysis is classical, it provides insights into the
behavior of quantum Kerr systems in the semiclassical regime where
photon numbers are large. Recent research has established
connections between classical Lyapunov exponents and quantum chaos
indicators such as out-of-time-ordered correlators (OTOCs)
\cite{Maldacena2016}.

The stable attractors we identified correspond to coherent states
in the quantum description, while the chaotic regions relate to
situations where quantum states exhibit rapid entanglement growth
and delocalization. This quantum-classical correspondence is
particularly relevant for superconducting circuit implementations,
which often operate in a mesoscopic regime where both classical
and quantum effects are important.

\subsection{Limitations and future work}

Several limitations of our current model should be acknowledged:
\begin{enumerate}
\item \textbf{Classical approximation:}
Our analysis is entirely classical, neglecting quantum effects
that may become significant at low field intensities or in
specialized configurations designed to enhance quantum
correlations.

\item \textbf{Simplified coupling:}
The coupling terms in our model represent instantaneous
interactions, neglecting potential time delays and
frequency-dependent effects that may occur in real optical
systems.

\item \textbf{Parameter restrictions:}
We have focused on symmetric configurations with identical
oscillator parameters $\epsilon_j$ to isolate the effects of coupling
topologies, but asymmetric parameters could reveal additional
interesting dynamics.
\end{enumerate}

Future work could address these limitations by:
\begin{itemize}
\item
Extending the model to include quantum effects, potentially
revealing connections to quantum chaos.

\item
Investigating asymmetric configurations with varied oscillator
parameters.

\item
Exploring the effects of time-delayed coupling, which could
introduce additional complexity and potential applications in
reservoir computing.

\item
Developing control strategies to stabilize desired attractors or
to switch between attractors for optical routing applications.
\end{itemize}

\section{Conclusions}

This paper has presented a comprehensive stability analysis of
three coupled Kerr oscillators in both triangular and sandwich
configurations, providing new insights into the dynamics of
coupled nonlinear optical systems with significant implications
for quantum computing. Through numerical simulations and Lyapunov
exponent analysis, we have characterized the system's behavior
across different parameter regimes, with particular focus on the
transition from regular to chaotic dynamics.

Our key findings can be summarized as follows:
\begin{enumerate}
\item \textbf{Multiple stable attractors:}
The subsystem consisting of the first oscillator exhibits three
distinct circular attractors with
different radii in phase space, dependent on initial conditions.
The complex basin structure of these attractors reveals the
intricate nature of the underlying dynamics. In quantum computing
implementations, these distinct states can serve as computational
basis states for information encoding.

\item \textbf{Configuration-dependent properties:}
The coupling configuration (triangular vs. sandwich) significantly
affects the attractor properties and system frequencies.
Counterintuitively, removing specific couplings in the sandwich
configurations leads to larger secondary and tertiary attractors
compared to the fully-coupled triangular arrangement. This finding
has important implications for quantum hardware design, indicating
that modifying coupling configurations can substantially enhance
computational performance.

\item \textbf{Damping-controlled transition to chaos:}
Lyapunov exponent analysis reveals a transition from stable to
chaotic dynamics as damping decreases. We identify critical
damping thresholds below which chaos emerges, with the undamped
system ($\gamma = 0$) exhibiting fully developed chaos marked by
positive Lyapunov exponents. Understanding these stability
characteristics is essential both for designing quantum operations
with predictable performance and for applications that
intentionally leverage chaos for computational advantage.

\item \textbf{Frequency-dependent stability windows:}
Even in chaotic regimes, certain pumping frequencies support
islands of stability, suggesting the possibility of controlling
the system's behavior through careful parameter selection. This
frequency dependence could be exploited for frequency-selective
quantum operations or for implementing multi-frequency encoding
schemes.

\item \textbf{Chaotic beats:}
The considered system can generate characteristic signals,
so-called chaotic beats. Unexpectedly, these chaotic beats were found in the case of strong coupling of the three Kerr oscillators---typically, this effect occurs in weakly coupled systems. The
multitude of parameters and the resulting richness of dynamical
behaviors suggest that this specific type of system dynamics can
emerge across a wide range of coupler parameter configurations.
\end{enumerate}

The significance of these results extends beyond the specific
system studied here. The mechanisms of transition to chaos that we
have identified---involving the competition between nonlinear
coupling and dissipation---are likely applicable to a wide range
of coupled nonlinear oscillator systems. Our findings on how
coupling topology affects stability may inform the design of
nonlinear optical devices where controlled chaos or switching
between multiple stable states is desired.

In the context of quantum computing, our work contributes to the
understanding of Kerr parametric oscillator systems that are
being increasingly utilized as fundamental building blocks in
quantum processors. The stability analysis we have presented
provides insights into parameter regimes suitable for implementing
high-fidelity quantum gates, error-resilient encoding schemes, and
specialized computing paradigms like quantum neural networks.

Potential applications of these results include optical switches
based on controlled transitions between attractors, secure
communications leveraging chaotic dynamics, random number
generation using the unpredictable nature of the chaotic regime,
multi-state optical memory elements utilizing the system's
multiple attractors, quantum gate implementations in
superconducting circuit platforms, and error correction schemes
exploiting the enhanced stability of specific parameter regimes.

Future work will focus on extending this analysis to asymmetric
configurations, including time-delayed coupling effects, and
developing experimental implementations to verify our theoretical
predictions. Additionally, exploring the quantum analogs of these
classical dynamics may reveal new phenomena at the
quantum-classical boundary, particularly in the context of quantum
chaos and its applications in quantum information processing.

\titleformat{\section}[block]{\filright\bfseries}{\thesection}{1em}{}
\section*{Acknowledgment}\affSettings
The authors would like to express their gratitude to Prof. Adam
Miranowicz for his valuable comments and careful reading of the
final version of the article. K.B. was supported by the Polish
National Science Centre (NCN) under the Maestro Grant No.
DEC-2019/34/A/ST2/00081.


\titleformat{\section}[block]{\filright\bfseries}{\thesection}{1em}{}

\end{multicols}

\end{document}